\documentclass[prl,twocolumn,showpacs,letterpaper]{revtex4}

\usepackage{graphicx}
\usepackage{dcolumn}

\newcolumntype{x}{D{.}{.}{0}}
\newcolumntype{y}{D{.}{.}{1}}
\newcolumntype{z}{D{.}{.}{2}}
\newcommand{\moc}{\multicolumn{1}{c}}

\newcommand{\mthc}{\multicolumn{3}{c}}
\newcommand{\ms}{\ensuremath{\mbox{m}/\mbox{s}^2}}

\begin{document}

\title{Test of the Equivalence Principle Using a Rotating Torsion Balance}
\date{\today}

\author{S.~Schlamminger}
\author{K.-Y.~Choi}
\author{T.A.~Wagner}
\author{J.H.~Gundlach}
\author{E.G.~Adelberger}
\affiliation{
Center for Experimental Nuclear Physics and Astrophysics, University of Washington, Seattle, WA, 98195}

\begin{abstract}
We used a continuously rotating torsion balance instrument to measure the acceleration difference of beryllium and titanium test bodies towards sources at a variety of distances. Our result $\Delta a_{N,Be-Ti}=(0.6\pm 3.1)\times 10^{-15}\;\ms$ improves limits on equivalence-principle violations with ranges from 1~m to $\infty$ by an order of magnitude. The E\"otv\"os parameter is $\eta_{Earth,Be-Ti}= (0.3 \pm 1.8)\times 10^{-13}.$  
By analyzing our data for accelerations towards the center of the Milky Way we find equal attractions of Be and Ti towards galactic dark matter, yielding $\eta_{DM,Be-Ti}= (-4  \pm 7 )\times 10^{-5}$.  Space-fixed differential accelerations in any direction are limited to less than $8.8\times 10^{-15}\;\ms$ with 95\% confidence.
\end{abstract}


\maketitle

The equivalence of gravitational mass and inertial mass is assumed as one of the most fundamental principles in nature. Practically every theoretical attempt to connect general relativity to the standard model allows for a violation of the equivalence principle\cite{Dam96}. Equivalence principle tests are therefore important tests of unification scale physics far beyond the reach of traditional particle physics experiments.
The puzzling discoveries of dark matter and dark energy provide strong motivation to extend tests of the equivalence principle to the highest precision possible.
 
Over the past two decades we have conducted laboratory tests of the equivalence principle\cite{Ade90,Su94,Bae99,Smi00}. This letter reports our latest and most precise measurement using a new, continuously-rotating torsion balance. The torsion balance compares the horizontal accelerations of test bodies made from two different materials. Acceleration differences that depend only on the test-body material violate the equivalence principle. We parameterize such equivalence-principle violating interactions by a Yukawa potential, which for two point objects is
\begin{equation}
V(r)=\alpha G 
(\frac{q}{\mu})_{1}
(\frac{q}{\mu})_{2}
\frac{m_1 m_2}{r_{12}}e^{-r_{12}/\lambda},
\label{eqn: yukawa}
\end{equation}
where the interaction range $\lambda=h/(m_{b}c)$ is given by the Compton wavelength of the presumed exchange boson of mass $m_{b}$, and which couples to the ``new charge'' $q$. The coupling strength $\alpha$ is measured in units of the Newtonian gravitational constant $G$, and $\mu$ represents the mass in atomic mass units.
The instrument consists of a highly sensitive torsion balance that is continuously rotated by a turntable.
The torsion balance is composed of a material-composition dipole pendulum suspended by a fine tungsten wire. Any  material-dependent acceleration on the composition dipole pendulum to objects not on the turntable produces a periodic twist of the torsion fiber. The twist angle, $\Theta$, was recorded using a corotating autocollimator. The phase and frequency of the sinusoidal variation of $\Theta$ allowed us to search for differential accelerations towards terrestrial and astronomical sources.  

\begin{figure} 
\includegraphics[width=8.6cm]{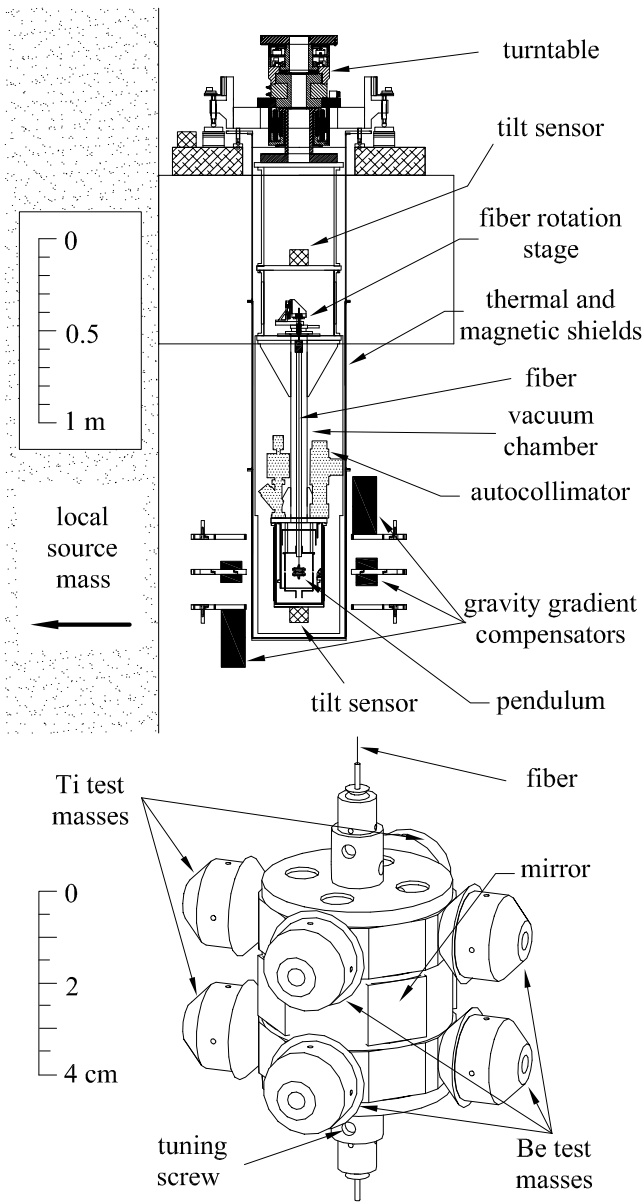}
\caption{Cross section of the apparatus (upper part). The entire  torsion balance  is suspended below a continuously rotating   turntable. Gravity gradient compensator masses were placed around the pendulum to reduce coupling to ambient gravitational gradients. The pendulum (lower part) carries four Ti and four Be masses in a   composition dipole.   
} 
\label{fig:schematics}
\vspace{-0cm}
\end{figure}

Figure~\ref{fig:schematics} shows a schematic drawing of the apparatus and the $70.3\;\mbox{g}$ pendulum. The pendulum body is a thin aluminum shell with fourfold azimuthal symmetry and up down reflection symmetry. It carries four beryllium and four titanium test masses in a horizontal dipole configuration. These two materials were chosen primarily to maximize the difference in baryon number ($B/\mu$ is $0.99868$ for Be and $1.001077$ for Ti), and secondly for experimental reasons, such as densities, magnetic properties and machinability. The Ti test bodies are hollow  to match the external shape and mass of the $4.84$~g Be test bodies to within $50$~$\mu$g. The test body shape allows us to reproducibly interchange the test bodies, to minimize alignment errors, and to equalize their gravitational interaction. 
The optical beam of the autocollimator is reflected from one of four mirrors located at the pendulum's midplane. The entire pendulum and all surfaces near the pendulum are plated with $\approx 300$~nm of gold.

The $1.07~\mbox{m}$ long torsion fiber is a $20\;\mu\mbox{m}$ thick tungsten wire with a torsion constant of $\kappa=2.34\;\times 10^{-9}\;\mbox{Nm}/\mbox{rad}$ yielding a free torsional oscillation period of  $T_{0}=798$~s and a quality factor $Q=5000\pm200$.  The top of the torsion fiber is attached to an eddy-current swing and vibration damper and tilt isolator. This assembly can be rotated with respect to the vacuum chamber by a small rotation stage. The vacuum is maintained at $\approx10^{-5}\;\mbox{Pa}$ by an ion pump.  The pendulum is surrounded by three layers of $\mu$-metal shielding on the rotating platform and one non-rotating $\mu$-metal shield. To reduce the sensitivity to temperature fluctuations and gradients a thermally-isolated copper tube inside the vacuum surrounds the fiber, the aluminum-vacuum chamber is surrounded by several layers of rotating and non-rotating thermal shields and insulation. The entire instrument is in a Styrofoam enclosure and the apparatus is located in a temperature-stabilized underground room. 

The turntable, below which the torsion balance is attached, is made from a custom-built aluminum air bearing~\cite{ProI}. The turntable rotation rate is controlled using feedback to an optical angle encoder with two read heads. The loop is realized by a digital signal processor controlling an eddy-current drive. The constancy of the rotation rate was limited by the angle encoder's linearity. Angle encoder imperfections were mapped by operating the torsion balance at rotation periods $T_{t}\ll T_0$ and the corrections were included in the feedback algorithm. The turntable rotation frequency was normally set to $\omega_{t}/(2\pi)= 2/3~T_{0}^{-1}=0.835$~mHz, where the signal-to-noise ratio was found to be optimal.
The pendulum's angular position and the signals of 31 sensors for temperature, tilt, etc. were recorded every 2.76~s.

Gravitational forces between our pendulum and local gravitational gradients can occur at the signal frequency. To minimize coupling to ambient gravitational field gradients, $Q_{l1}$, the pendulum was highly symmetrical, with the nominal mass moments, $q_{l1}$, vanishing for $l<7$. 
In addition, the pendulum's $q_{20}$, $q_{30}$ and $q_{40}$ moments were designed to be zero to avoid gravitational coupling due to a small misalignments of the pendulum \cite{Su94}. 
The ambient $Q_{21}$ and $Q_{31}$ fields at the pendulum's position were compensated with $888~\mbox{kg}$ of lead and $8.8~\mbox{kg}$ of aluminum, respectively.  Lead bricks placed about 1~m from the pendulum were used to null the remnant $Q_{21}$-field. The gradient fields were measured with a special gradiometer pendulum that was configured to have either a large $q_{21}$ or $q_{31}$ moment. We found that the $Q_{21}$-field varied by as much as $\pm$1\% during a year, which we attribute to changes of the water table. Once the equivalence principle pendulum was installed, its residual $q_{21}$- and $q_{31}$-moments were measured by rotating the $Q_{21}$- and $Q_{31}$-compensators by 180$^{\circ}$, doubling the uncompensated fields. The pendulum's unwanted mass moments were then minimized by adjusting four screws on the pendulum body until only a small gravitational coupling remained (see Table~\ref{tab1}), which was later subtracted from the data.  

The turntable must rotate about local vertical since a tilt of the attachment point of the torsion fiber causes a small apparent rotation of the pendulum. The tilt of the turntable was continuously measured with level sensors on the rotating platform. The once-per-revolution component of the tilt was minimized by a feedback loop that changed the length of two of
the turntable's support legs by controlling their temperature with Peltier elements.
This system nulled the periodic tilt of the level sensor to within $\pm$3 nrad.
At the pendulum body, $1.7$~m below the feedback sensor, local vertical was different by 55~nrad and a correction to the data was required. We inferred this tilt by using a second tilt sensor $0.2$~m below the pendulum. We also found the change in local vertical consistent with our local mass integration. The data were corrected using the tilt at the pendulum and a tilt matrix characterizing the tilt sensitivity. The tilt matrix was measured for all four pendulum mirrors by deliberately tilting the turntable rotation axis. The magnitude of the  tilt matrix  ranged from $\approx$~1\% to $\approx$~7\% depending on the mirror.

The largest entry in our error budget arises from systematic and statistical uncertainties associated with temperature gradients and fluctuations. We assessed the sensitivity to temperature gradients by placing temperature-controlled panels on opposite sides of the apparatus to produce exaggerated thermal gradients. The quoted uncertainty is limited by sensor noise in resolving the small temperature gradients that occurred during equivalence principle data taking.  
By applying a magnetic field and field gradients using coils, as well as strong permanent magnets, we found the magnetic coupling to be small (Table \ref{tab1}).

\begin{figure}[htb]
  \includegraphics[width=8.0cm]{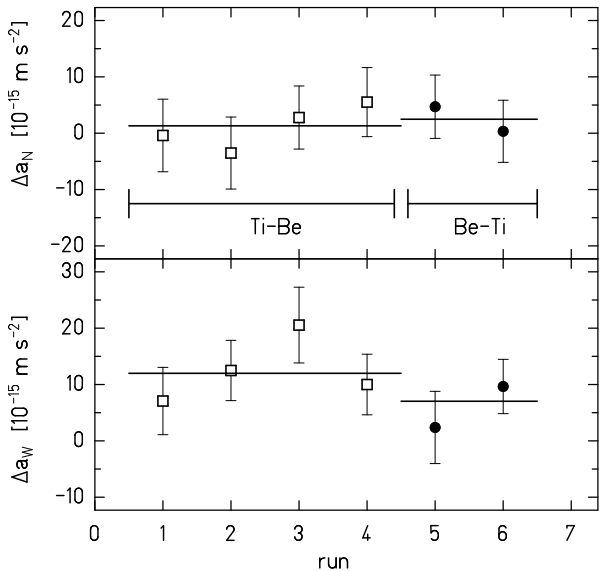}
\caption{
Shown are measured differential accelerations towards North (top) and West. After the first four data runs the Be and Ti test bodies were interchanged on the pendulum frame. A violation of the equivalence principle would appear as a difference in the means (lines) of the two data sets. The offset acceleration is due to systematic effects that  follow the pendulum frame but not the composition dipole. The data have been corrected for tilt and gravity gradients, but only the statistical uncertainties are shown.}
\label{fig:rawdata}
\vspace{-0cm}
\end{figure}

\begin{table}[htb]
\caption{The raw differential accelerations between Be and Ti towards North (N)and West (W) are shown in line 1. Lines 2 to 5 list corrections that were applied and the bottom line gives our corrected results. Uncertainties are 1$\sigma$.}
\begin{ruledtabular}
\begin{tabular}{lycyycy}
\moc{differential acceleration in}           & 
\mthc{$\Delta a_{N,Be-Ti}$} & \mthc{$\Delta a_{W,Be-Ti}$} \\
                                & \mthc{$(10^{-15}\;\ms)$}&
                                  \mthc{$(10^{-15}\;\ms)$}\\
\hline
as measured (statistical)
&3.3  & {$\;\pm$} & 2.5      &-2.4   & {$\;\pm$} & 2.4  \\
\hline
residual gravity gradients        
&1.6  & {$\;\pm$} & 0.2      & 0.3  & {$\;\pm$} & 1.7 \\
tilt                     
&1.2  & {$\;\pm$} & 0.6      & -0.2  & {$\;\pm$} & 0.7 \\
magnetic
&0  & {$\;\pm$} & 0.3      & 0 & {$\;\pm$} & 0.3 \\
temperature gradients
&0  & {$\;\pm$} & 1.7      & 0 & {$\;\pm$} & 1.7 \\
\hline
corrected                       
&0.6& {$\;\pm$} & 3.1       &-2.5  & {$\;\pm$} & 3.5  \\
\end{tabular}
\end{ruledtabular}
\label{tab1}
\end{table}

We collected 75 days of equivalence-principle data in two sets. The orientation of the pendulum with respect to the vacuum chamber was changed by $180^{\circ}$ once a day, mostly to cancel the effect of small turntable rotation rate imperfections. Approximately biweekly we changed the orientation of the pendulum by $90^{\circ}$ and took data using opposite mirrors. Between the two sets, we interchanged the test bodies with those on the other side of the pendulum, reversing the composition dipole on the pendulum frame to eliminate spurious signals associated with the torsion fiber, the pendulum body, or the magnetic damper. An equivalence principle violation would cause a difference in the measured acceleration of the two sets  (Fig.~\ref{fig:rawdata}). 

Our data analysis began by using a digital notch filter to remove the pendulum's free torsional oscillation. The data were then divided into segments containing 2 turntable revolutions ($\approx 2400$~s). Nine  harmonics of the turntable rotation frequency, an offset, a linear and a quadratic drift were fitted to the data in each segment. About 7\% of the segments were excluded from the analysis due to spikes in the ion pump current or due to abrupt changes of the turntable axis from local vertical. 
The coefficients of the $\sin{(\omega_{t} t)}$ and $\cos{(\omega_{t} t)}$ terms were used to extract the signals of interest and their scatter was used to determine the statistical uncertainty.
The pendulum data were corrected in amplitude and phase for the pendulum's dynamic response to external torques and for electronic attenuation and time delays. After correcting for systematic effects (Table \ref{tab1}), the  data were reduced to material-dependent differential acceleration components along North and West directions. Figure~\ref{fig:rawdata} shows these components for six data runs. We observed a small unexplained offset in the East-West direction that followed the suspension system (pendulum body, torsion fiber, swing damper) but the signals tracking the composition dipole were unresolved in all directions. 

The composition dependent accelerations are 
\begin{eqnarray*}
a_{N}(\mbox{Be})-a_{N}(\mbox{Ti}) &=& (0.6\pm 3.1)\times 10^{-15}\;\mbox{m}/\mbox{s}^{2}
\;\mbox{and}\\
a_{W}(\mbox{Be})-a_{W}(\mbox{Ti})&=& (-2.5\pm 3.5)\times 10^{-15}\;\mbox{m}/\mbox{s}^{2}
\label{eqn:results}
\end{eqnarray*}
with titanium being more attracted to the South and to the West. Our best limit on the classical equivalence principle
parameter\cite{Wil93} $\eta$ is
\begin{equation}
\eta(\mbox{Be}-\mbox{Ti})= \frac{\Delta a_{N}}{a_{\perp}^g}= (0.3\pm 1.8)\times 10^{-13}.
\label{eqn:eta}
\end{equation}

\begin{figure}[htb]
\includegraphics[width=8.6cm]{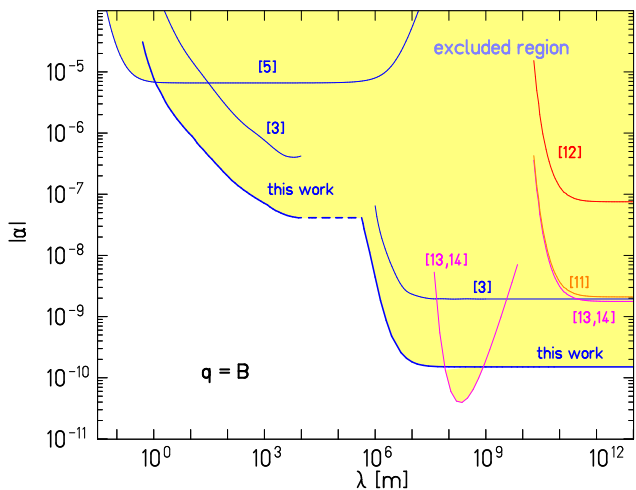}
\caption{New upper limits on Yukawa interactions coupled to baryon number with 95\% confidence. The uncertainties in  the source integration is not included in this plot. The numbers indicate references. The shaded region is experimentally excluded. Preliminary models for $10\;\mbox{km} < \lambda < 1000\;\mbox{km}$ indicate that the limit on $\alpha$ is smaller than the dashed line.}
\label{fig:results}
\vspace{-0cm}
\end{figure}

Figure~\ref{fig:results} shows the limits on the strength, $\alpha$, of a new
interaction (Eq.~1) as a function of range $\lambda$. To establish these limits we used the mass density and charge content of the environment surrounding the torsion balance to create a source model. For $\lambda= 1-100$~m the source is dominated by a hill sloping towards the East. For $\lambda<10\;$km the local topography and bedrock become significant. At ranges between 10~km and 1000~km, preliminary results using large scale density and composition models indicate that the limit on $\alpha$ is better than the dashed line shown in Fig.~\ref{fig:results}. A detailed description of the model and limits will be included in a future publication. We used an elliptical layered Earth model\cite{Su94, Dzi81, Mor80} for $\lambda>1000\;$km. For this range the source mass is located towards the North.      

\begin{figure}[htb]
\includegraphics[width=8.0cm]{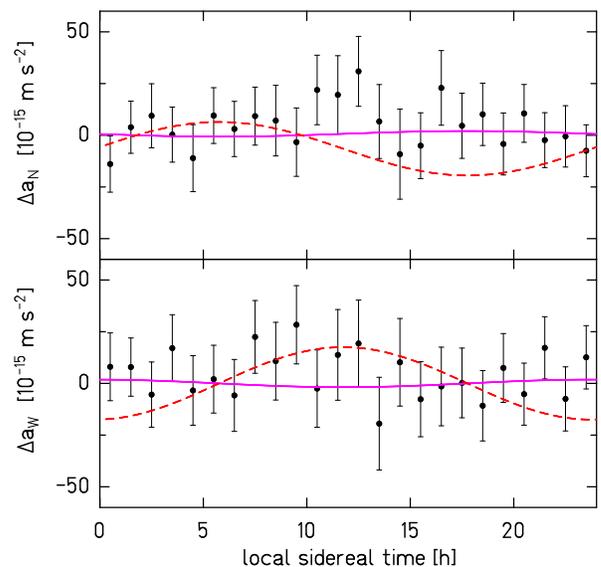}
\caption{The averaged differential acceleration of Be and Ti towards North and West as a function of sidereal time. The dashed line represents a hypothetical signal of $20\times 10^{-15} \;   \mbox{m}/\mbox{s}^2$. The solid line is the best fit toward the galactic center ($\Delta   a=(-2.1 \pm 3.1)\times 10^{-15}\;\ms$). }
\label{fig:sunfit}
\vspace{-0cm}
\end{figure}

Equivalence-principle violating interactions associated with an
astronomical source are additionally modulated by a solar or sidereal
frequency: 
\newcommand{\dph}{(\phi-\phi_0)}
\begin{eqnarray*}
\Delta a_{N}\!\!&\!\!=&\!\!\cos\theta  \big( - \Delta a  \cos{\dph} -
\Delta \tilde{a} \sin{\dph} \big) + \!o_{N} , \\
\Delta a_{W}\!\!&\!\!=&\!\!\cos\theta  \big(   \Delta a  \sin{\dph} - \Delta \tilde{a} \cos{\dph} \big)+\!o_{W},
\end{eqnarray*}
with $\theta$ and $\phi$ being the altitude and azimuth of the astronomical source,  $\Delta a$  the differential acceleration towards the source and $\Delta \tilde{a}$ its quadrature component; $o_{N}$ and $o_{W}$ are possible instrument offsets.  Figure~\ref{fig:sunfit} shows the averaged $\Delta a_{N}$ and $\Delta a_{W}$ versus sidereal time. A simultaneous fit of $\Delta a_{N}$ and $\Delta a_{W}$ towards the galactic center yields
\begin{eqnarray*}
a(Be)-a(Ti)=\Delta a&=& ~(-2.1\pm 3.1)\times 10^{-15}\;\mbox{m}/\mbox{s}^{2}\;,\\
\tilde{a}(Be)-\tilde{a}(Ti)= \Delta \tilde{a}&=&(2.7\pm 3.1)\times 10^{-15}\;\mbox{m}/\mbox{s}^{2}.
\label{eqn:galresults}
\end{eqnarray*}
Since only about a quarter of the total acceleration of the solar system towards the center of our galaxy  
is caused by galactic dark matter~\cite{Stu93}, we find $\eta_{DM,Be-Ti}= (-4  \pm 7 )\times 10^{-5}$.

With 95\% confidence we constrain space-fixed differential accelerations in any direction to be smaller than  $\Delta a= 8.8\times 10^{-15}\;\mbox{m}/\mbox{s}^{2}$.

We have substantially improved the limits on the strength of an equivalence-principle violating, long-range interaction.
We are currently broadening our search by using other test-body materials and improving the sensitivity of our torsion balance. 

\nocite{Bra72,Rol64,Smi00,Wil04,Tal88}
This work was supported by NSF Grants PHY0355012, PHY0653863 and by NASA Grant NNC04GB03G, and DOE funding for the CENPA laboratory. Blayne Heckel, Stephen Merkowitz, Erik Swanson, Phil Williams, Ulrich Schmidt, Tom Butler and Chris Spitzer have contributed to the apparatus development.

\end{document}